\documentstyle[preprint,aps,epsf]{revtex}
\begin{document}
\draft
\title{Modeling Morphology of Cities and Towns }

\author{Hern\'an A. Makse$^1$, Shlomo Havlin$^{1,2}$, and H. Eugene
Stanley$^1$}

\address{$^1$Center for Polymer Studies and Dept. of Physics,
 Boston University, Boston, Massachusetts 02215 }

\address{$^2$Department of Physics, Bar-Ilan University,
Ramat-Gan, ISRAEL}

\date{\today}

\maketitle

\bigskip\bigskip

{\bf Predicting urban growth is important for practical reasons,
and
also for the challenge it presents to theoretical frameworks for cluster
dynamics \cite{fractal,benguigui1,benguigui2}.
Recently, the model of diffusion limited aggregation
(DLA)\cite{dla,vicsek} has been applied to describe urban growth
\cite{fractal}, and results in tree-like dendritic structures which have
a core or ``central business district'' (CBD).  The DLA model predicts
that there exists only one large fractal cluster that is almost perfectly
screened from incoming ``development units'' (people, capital,
resources, etc), so that almost all the cluster growth occurs in the
extreme peripheral tips.
Here we propose and
develop an alternative model to DLA that describes the morphology and
the area distribution of systems of cities, as well as the
scaling of the urban perimeter
of individual cities.
Our results agree both  qualitatively and quantitatively with actual urban
data.   The resulting growth morphology
can be understood in terms of the effects of interactions among the
constituent units forming a urban region, and can be modeled using the
correlated percolation model in the presence of a gradient.  }

In the model we now develop
we take into account the following points:

$(i)$ Urban data on the population density $\rho(r)$ of actual urban systems
are known to conform to the relation
 \cite{clark},
$\rho(r) = e^{-\lambda r}$, where $r$ is the radial distance from the
CBD situated at the core, and $\lambda$ is the density gradient.
Therefore, in our model
the development units are positioned
with an occupancy probability $p(r) \equiv \rho(r)$ that behaves
in the same fashion as is known experimentally.

$(ii)$ We also take into account the
fact that in actual urban systems,
the development units are not
positioned at {\it random}.
Rather, there exist {\it correlations} arising from the fact that when a
development unit is located in a given place, the probability of adjacent
development units increases naturally--- i.e.,
each site is not independently occupied by
a development unit, but is occupied with a
probability that depends on the occupancy of the
neighborhood.

In order to quantify these ideas, we
consider the {\it correlated} percolation model
\cite{coniglio,weinrib,weinrib2,sona}.
In the limit where correlations are so small as to be negligible
\cite{percolation}, a site at position
{\bf\it r} is occupied if the occupancy variable $u$({\bf\it r}) is
smaller than the occupation probability $p$({\bf\it r}); the variables
$u$({\bf\it r}) are uncorrelated random numbers.  To introduce
correlation among the variables, we convolute the uncorrelated variables
$u$({\bf\it r}) with a kernel $G(${\bf\it r}), and define a new set of
random numbers $\eta$({\bf\it r})$\equiv \int G$({\bf\it
r-r'})$u$({\bf\it r'})$d${\bf \it r}'.  The numbers $\eta$({\bf \it r})
have long-range power-law correlations that decay as $r^{-\alpha}$ if,
e.g., we
choose  $G(${\bf \it r}$) \equiv
1/(1+r^2)^{\alpha/4}$, where $r\equiv|${\bf\it r}$|$.
This choice assures that the
correlations extend to the entire lattice.
The Fourier transform of $G(r)$, needed in order to
convolute the variables in the Fourier space, is $G(q) =
q^\beta~ K_\beta(q)$, where $K_\beta(q)$ is the modified Bessel
function of order $\beta=(\alpha-2)/4$.
The assumption of power-law interactions is motivated by the
fact that the ``decision'' for a development unit to be placed in a
given location decays gradually with
the distance from an occupied neighborhood.
The correlations have the
effect of agglomerating the units around a urban area, and they
also quantify the interactions between clusters that are not physically
connected. The correlation exponent $\alpha$ is the only parameter to be
determined by empirical observations.

To  discuss  the morphology of a system of cities generated in the
present model,
we show in
Fig. \ref{static}
our simulations of  correlated urban systems
for a fixed value of the density gradient
$\lambda$,
and for different degree of correlations quantified by the exponents $\alpha$.
We note the
effect of correlations among the clusters.
The strongly correlated system
of Fig. \ref{static}$a$ compares
better with actual urban
settlements (i.e. see Fig. \ref{dynamics}$a$) than the weakly correlated
system of Fig. \ref{static}$b$.
In the simulated systems the largest
city is situated in the core, which is regarded as the attractive center
of the city,
and is surrounded by small towns.
All towns are nearly
compact near their centers and become less compact near
their  boundaries, in qualitative agreement with empirical data on
actual large  cities such as Berlin, Paris, London, etc. (see i.e. Refs.
\cite{fractal,berlin}).

Next we
test the proposed  model quantitatively.
The urban boundary of the largest city
is defined to be the perimeter of the cluster connected to the CBD.
Since $p(r)$ decreases as one moves away from the core, the probability that
the largest cluster remains connected decreases with $r$.
Hence $r_f$, the {\it mean} distance of the perimeter  from the center,
is determined by the value of $r$ for which $p(r)$ equals the
percolation threshold---i.e.
$p(r_f)=p_c$, so
$r_f = \lambda^{-1}\ln(1/p_c)$ \cite{sapoval,sapoval2}.

The urban
boundary in the model   has the scaling properties of the  external
perimeter of a correlated percolation cluster in the presence of a
gradient, which we calculate to
have
fractal dimension
$D_e \simeq 1.3$ for the uncorrelated
case, and $D_e\simeq1.4$ for
strong
correlations $(\alpha\to 0)$. These values are
consistent with actual urban data,
for which values of $D_e$ between $1.2$ and $1.4$
are measured \cite{fractal}.
Near the  frontier and on length scales smaller than the width of the
frontier, the largest cluster has fractal dimension $d_f \simeq 1.9$.
In contrast to the value of
$D_e$, we find that the value of $d_f$ is not affected by the
long-range correlations parametrized by $\alpha$.

So far,
we have shown how the correlations between the occupancy probabilities
can account for the irregular
morphology of the towns in a urban system.
As can be seen in  Fig. \ref{dynamics}$a$,
the towns surrounding a large city like Berlin are characterized
with a wide range of sizes.
We are interested in the laws
that quantify the distribution $N(A)$ of the area of the towns
in such a  system. Here, $A$ is the area occupied by a given town.
To this end, we calculate the actual
distribution of the areas of the urban settlements around Berlin and
London.
We first digitize the empirical data of Fig. 4.1 of
Refs. \cite{berlin} (which are shown in Fig. \ref{dynamics}$a$), and
Fig. 10.8 of Ref. \cite{fractal}. Then, we
count  the number  of towns with area $A$, putting the result in
logarithmicaly spaced bins, and averaging over the size of the bin.
The results for the distributions $N(A)$ calculated in this way
are shown in Fig. \ref{nu}$a$, where we see
that for both Berlin and
London,
$N(A)$ follows a power-law.

This new result of a power law area distribution of towns, $N(A)$,
can be understood in the context of our model.
Insight into this
distribution
can be developed by first noting that
the small ``towns'' surrounding the largest city
are all situated at distances $r$ from the CBD
such that $p(r) < p_c$ or $r > r_f$. Therefore, we find $N(A)$,
the cumulative area distribution of
clusters of area $A$, to be
\begin{equation}
\label{distribution}
N(A) \equiv \int_0^{p_c} n(A,p) ~ dp \sim A^{-(\tau+1/d_f \nu)}.
\end{equation}
Here, $n(A,p) \sim A^{-\tau}g(A/A_o)$ is defined
to be the average number of clusters containing $A$ sites for a given
$p$ {\it at a fixed
distance} $r$, and
$\tau=1+2/d_f$.
Here,
$A_o(r)\sim |p(r)-p_c|^{-d_f\nu}$  corresponds to the
maximum  typical area
of a ``town'' situated  at a distance $r$ from the CBD,
while $g(A/A_o)$ is a
scaling function that decays rapidly (exponentially) for $A>A_o$.
The exponent $\nu=\nu(\alpha)$ is
defined by $\xi(r) \sim |p(r)-p_c|^{-\nu}$,
where $\xi(r)$ is the connectedness length that
represents the mean
linear extension of a cluster at a distance $r$ from the CBD.

In our numerical simulations we
find a drastic increase of
$\nu(\alpha)$ with the increase of the long-range correlations
$(\alpha\to 0)$ (Fig.  \ref{nu}$b$) \cite{sona}.
The exponent $\nu(\alpha)$ affects the
area distribution of the small clusters around the CBD (Fig.
\ref{nu}$c$), as specified by Eq.(\ref{distribution}), and can be used
to quantify the degree of interaction between the CBD and the small
surrounding towns. For instance, for a strongly correlated system of
cities characterized by small $\alpha$, $\nu(\alpha)$ is large so that
the area $A_o(r)$ and the linear extension $\xi(r)$ of the towns will be
large even for towns situated away from the CBD. This effect is
observed in the correlated systems of cities of Fig.
\ref{static}.

In Fig. \ref{nu}$a$ we also plot the power-law for the area distribution
predicted by Eq. (\ref{distribution}) along with the real data for Berlin
and London.
We find that the slopes of the plots for both cities
are consistent with the prediction (dashed line) for
the case of highly correlated systems.
These results quantify the qualitative  agreement
between the morphology of actual urban areas and the
strongly correlated urban systems obtained in our simulations.

Before concluding,
we discuss a generalization of our static model
to describe the dynamics of urban growth. Following the
pioneering work of Clark \cite{clark},
the density gradient  $\lambda(t)$ was calculated as a function of the time
for many urban areas around the world \cite{mills}.
Empirical studies show that  the population density profile of cities
presents  a remarkable pattern of decline with time, which is
quantified by the decrease of
$\lambda(t)$ with time (see Fig. \ref{berdensity}
and Table 4 in Ref. \cite{mills}) .
In the context of our model, this flattening pattern can be explained
as follows.
The model of  percolation in a gradient
can be related to a dynamical model
of units  diffusing from a central seed or core \cite{sapoval,sapoval2}.
In this dynamical system, the units are allowed to diffuse on a
two-dimensional lattice by hopping to nearest-neighbor positions. The
density of units at the core remains constant: whenever a unit diffuses
away from the core, it is replaced by a new unit. A well-defined
diffusion front, defined as the boundary of the cluster
of units that is
linked to the central core, evolves in time.
The diffusion front corresponds to the urban boundary of the central city.
The static properties of the
diffusion front of this system
were found to be the same as those predicted by
the gradient percolation model
\cite{sapoval,sapoval2}. Moreover, the dynamical model can explain
the decrease of
$\lambda(t)$ with time observed empirically.
As the
diffusion front situated around $r_f$
moves away from the core,
the city grows and the density gradient decreases since
$\lambda(t) \propto 1/r_f$. Thus, the dynamics in the model are
quantified by a decreasing  $\lambda(t)$, as occurs in actual urban areas.
These considerations are tested in Fig. \ref{dynamics}$b$, which
shows our dynamical urban simulations of a strongly
interacting system of cities characterized by a correlation exponent
$\alpha = 0.05$ for three different values of $\lambda$.
Qualitative agreement is observed between the morphology of the
cities and towns of the actual data of Fig. \ref{dynamics}$a$ and the
simulations of Fig. \ref{dynamics}$b$.

\begin{figure}
\narrowtext
\caption{Simulations of urban systems
for different degree of correlations.
Here, the urban areas  are red,
and the external perimeter or urban
boundary of the largest cluster connected to the CBD is light green. In
all the figures, we fix the value of the density gradient to be
$\lambda=0.009$.
$(a)$ and $(b)$ Two different examples of interactive systems of cities
for correlation exponents $\alpha=0.6$ and  $ \alpha=1.4$, respectively.
The development
units are positioned with a probability that decays exponentially with
the distance from the core. The units are located not randomly as in
percolation, but
rather in a correlated fashion depending of the neighboring occupied areas.
The correlations are
parametrized by the exponent $\alpha$. The strongly correlated case
corresponds to small $\alpha$ ($\alpha \to 0$). When
$\alpha>d$, where $d$ is the spatial dimension of the substrate
lattice ($d=2$ in our case),
we recover the
uncorrelated case.
Notice
the tendency to more compact clusters as we increase the degree of
correlations ($\alpha \to 0$).
$(c)$
As a zeroth order approximation, one might imagine
the morphology predicted in the extreme limit whereby
development units are positioned  at {\it random}, rather than in the
correlated  way of Figs. $1a$ and $1b$. The results for this
crude approximation of a non-interactive (uncorrelated) system of cities
clearly display a drastically different morphology than
found from data on real cities (such as shown in Fig. $2a$).
The non-interactive limit looks unrealistic in comparison with
real cities, for the
lack of interactions creates a urban area characterized by many small
towns spread loosely around the core.}
\label{static}
\end{figure}

\begin{figure}
\narrowtext
\caption{Qualitative comparison between the actual urban data and the
proposed model. $(a)$ Three steps of the growth with time of Berlin and
surrounding towns
(from Ref. [$13$]). Data are shown for the years 1875,
1920, and 1945 (from top to bottom). The flattening of the
density gradient is evident  and
corresponds to the decentralization of the urban area as the city grows.
$(b)$ Dynamical urban simulations of the proposed model.
We fix the  value of the
correlation exponent to be $\alpha = 0.05$ (strongly correlated case),
and choose the occupancy probability $p(r)$
to correspond to  the data of
Fig. $2a$.}
\label{dynamics}
\end{figure}

\begin{figure}
\narrowtext
\caption{
$(a)$ Log-log plot of the area distribution $N(A)$
of the actual towns around Berlin and
London. In the case of Berlin, the  data are
shown for the years 1920
and 1945 (corresponding to the last two panels
in Fig. 2$a$), while the data of
London are for the year 1981. A
power-law  is observed for the area distributions of both urban systems.
The dotted line shows the predictions of our model for
the uncorrelated case (slope$=2.45$),
while the dashed line gives results for the
strongly correlated case (slope$=2.06$). Note that
the area distributions for both cities agree much better
with the strongly correlated case. The images have been
digitized
using an Apple Scanner, of resolution 150 dots per inche,
and in this figure $A$ denotes  the number of
sites covered by the development units for a given town.
$(b)$ Correlation length exponent $\nu(\alpha)$ as a function of the
correlation exponent $\alpha$. Notice the drastic variation of $\nu$
with the increase of the correlations $(\alpha \to 0)$.  This, in turn,
corresponds to more compact clusters as can be observed in Fig. 1. $(c)$
Log-log plot of the
area distribution $N(A)$ calculated for the present model for
different degrees of correlation. From top to bottom, $\alpha=0.2$,
$\alpha=0.8$, $\alpha=1.4$, and uncorrelated case.
The linear fits correspond to the predictions of Eq. ($1$) using the
values of $\nu(\alpha)$ from Fig. 3$b$, and $d_f=1.9$.}
\label{nu}
\end{figure}

\begin{figure}
\narrowtext
\caption{Semi-log plot of the
density profile $\rho(r)$ for the three different stages in the growth of
Berlin shown in Fig. $2a$, corresponding to the years
 1875, 1920, and 1945 from bottom to top. Least square
fits yield the results
$\lambda \simeq 0.030$,
$\lambda \simeq 0.012$, and $\lambda \simeq 0.009$, respectively,
showing the decrease of $\lambda$ with time.
We used these density profiles in the
dynamical simulations of Fig. $2b$.}
\label{berdensity}
\end{figure}

\end{document}